\documentclass[conference]{IEEEtran}

\usepackage{algorithm}
\usepackage{flushend}
\usepackage{algorithmic}
\usepackage{amsfonts}
\usepackage{amsmath}
\usepackage{amssymb}
\usepackage{array}
\usepackage{booktabs}
\usepackage{comment}
\usepackage{enumerate}
\usepackage{enumitem}
\usepackage{float}
\usepackage{graphics}
\usepackage{graphicx}
\usepackage{lineno}
\usepackage{multirow}
\usepackage{mathtools}
\usepackage{threeparttable}
\usepackage{soul}
\usepackage{url}
\usepackage{balance}
\usepackage[bookmarks=true,breaklinks=true,colorlinks,linkcolor=black,citecolor=blue,urlcolor=black]{hyperref}

\usepackage{textcomp}
\usepackage{xcolor}

\begin{document}

\title{FeatureBox: Feature Engineering on GPUs\\ for Massive-Scale Ads Systems}

\author{\IEEEauthorblockN{Weijie Zhao, Xuewu Jiao, Xinsheng Luo, Jingxue Li, Belhal Karimi, Ping Li}\\
\IEEEauthorblockA{
Cognitive Computing Lab, Baidu Research \\
Baidu Search Ads (Phoenix Nest), Baidu Inc.\\
10900 NE 8th St. Bellevue, Washington 98004, USA \\
No. 10 Xibeiwang East Road, Beijing 100193, China\\\\
\{weijiezhao, jiaoxuewu, luoxinsheng, lijingxue01, belhalkarimi, liping11\}@baidu.com}
}

\maketitle


\begin{abstract}
Deep learning has been widely deployed for online ads systems to predict Click-Through Rate (CTR). Machine learning researchers and practitioners frequently retrain CTR models to test their new extracted features. However, the CTR model training often relies on a large number of raw input data logs. Hence, the feature extraction can take a significant proportion of the training time for an industrial-level CTR model. In this paper, we propose FeatureBox, a novel end-to-end training framework that pipelines the feature extraction and the training on GPU servers to save the intermediate I/O of the feature extraction. We rewrite computation-intensive feature extraction operators as GPU operators and leave the memory-intensive operator on CPUs. We introduce a layer-wise operator scheduling algorithm to schedule these heterogeneous operators. We present a light-weight GPU memory management algorithm that supports dynamic GPU memory allocation with minimal overhead. We experimentally evaluate FeatureBox and compare it with the previous in-production feature extraction framework on two real-world ads applications. The results confirm the effectiveness of our proposed method.\\
\end{abstract}

\begin{IEEEkeywords}
CTR Prediction; GPU; Large-Scale Machine Learning Framework
\end{IEEEkeywords}

\vspace{0.25in}
\section{Introduction}\label{sec:intro}
\vspace{0.1in}

Deep learning has been widely employed in many real-world applications, e.g., computer vision~\cite{goodfellow2014generative,he2016deep,voulodimos2018deep,dosovitskiy2021image}, data mining~\cite{fei2019hierarchical,nguyen2019machine,kang2020novel,li2021learning,merhej2021what,zerveas2021transformer}, and recommendation systems~\cite{covington2016deep,cheng2016wide,wei2017collaborative,zhang2019deep,ma2020temporal,li2020video,zamany2022towards}.  In recent years, sponsored online advertising also adopts deep learning techniques to predict the Click-Through Rate (CTR)~\cite{fan2019mobius,zhao2019aibox, liu2020autogroup,zhu2020emsembled,fei2021gemnn,qi2021pp,sheng2021one,zhao2021rlnf,zhang2021multi,yu2022boost}. Unlike common machine learning applications, the accuracy of the CTR prediction is critical to the revenue.
In the context of a many-billion-dollar online ads industry, even a $0.1\%$ accuracy increase will result in a noticeable revenue gain~\cite{zhao2020distributed}. In this work,
we identify two major paths to improve the model accuracy. The first area is to propose different and enhanced model architectures. Every improvement in this direction is considered a fundamental milestone in the deep learning community---and does not happen often in the CTR prediction industry.
The other (more practical) is feature engineering, i.e., to propose and extract new features from the raw training data. The benefit of feature engineering is usually neglected in common deep learning applications because of the general belief that deep neural networks inherently extract the features through their hidden layers. However, recall that CTR prediction applications are accuracy-critical, hence, the gain from an improved feature engineering strategy remains attractive for in-production CTR prediction models.
Therefore, in order to achieve a better prediction performance, CTR deep learning models in real-world ads applications tend to utilize larger models and more features extracted from raw data logs.

Testing on the historical and online data is the rule-of-the-thumb way to determine whether a new feature is beneficial. Every new feature with positive accuracy improvement (e.g., $0.1\%$) is included into the CTR model. Machine learning researchers and practitioners keep this feature engineering trial-and-error on top of the current in-production CTR model. As a result, the in-production CTR model becomes larger and larger with more and more features. To support the trial-and-error research for new features, it requires us to efficiently train massive-scale models with massive-scale raw training data in a timely manner. Previous studies~\cite{zhao2020distributed} propose hierarchical GPU parameter server that trains the out-of-memory model with GPU servers to accelerate the training with GPUs and SSDs. With a small number of GPU servers, e.g., 4, can obtain the same training efficiency as a CPU-only cluster with hundreds of nodes. The training framework focuses on the training stage and assumes the training data are well-prepared---the training data are accessed from a distributed file system.

However, preparing the training data is not trivial for industrial level CTR prediction models---with $\sim$ $10^{12}$ features. The feature extraction from raw data logs can take a significant proportion of the training time. In addition to the frequent retraining for new feature engineering trials, online ads systems have to digest a colossal amount of newly incoming data to keep the model up-to-date with the optimal performance. For the rapid training demands, optimizing the feature extraction stage becomes one of the most desirable goals of online ads systems.
This latter point is the scope of our contribution.

\begin{figure*}[htbp]
\centering
\includegraphics[width=4.4in]{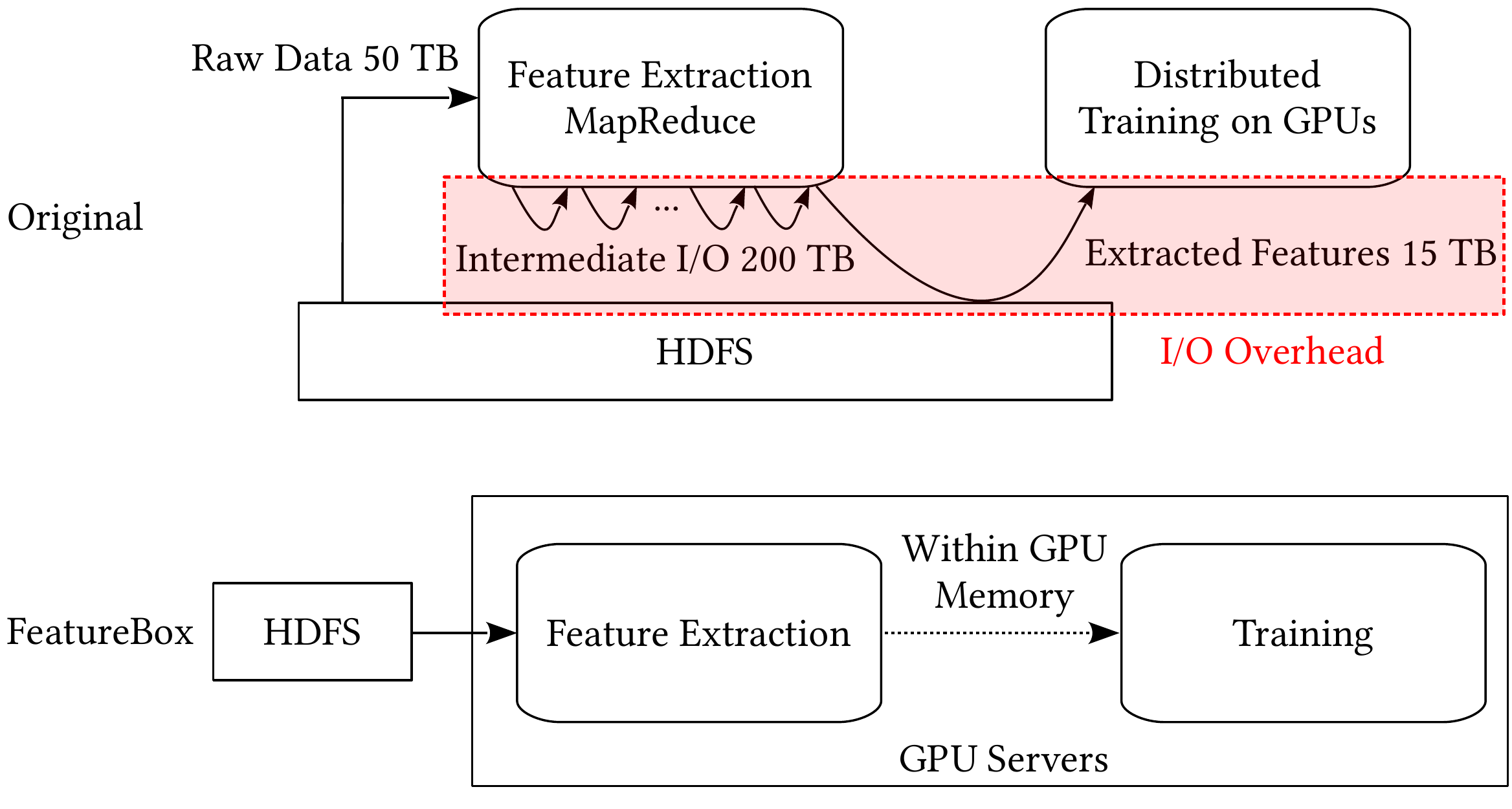}

\caption{A visual illustration for the original feature extraction and training workflow (upper); and our proposed FeatureBox (lower).}\label{fig:feabox}\vspace{0.1in}
\end{figure*}

\newpage

\noindent\textbf{Training workflow.}
The upper part of Figure~\ref{fig:feabox} depicts a visual illustration of the feature extraction. Due to the large amount of raw data, the original feature extraction task is constructed as MapReduce~\cite{dean2008mapreduce} jobs that compute feature combinations, extract keywords with language models, etc.
Those MapReduce jobs frequently read and write intermediate files with the distributed file system (i.e., HDFS~\cite{borthakur2008hdfs}). The intermediate I/O can be as large as 200 TB. Once the features are extracted, we also need to materialize them to the $\sim$$15$ TB extracted features to the HDFS so that the following distributed training framework can read them from the distributed file system.
This training workflow incurs rapid communication with HDFS that generates heavy I/O overhead.

One straightforward question can be raised: \textit{Can we perform the feature extraction within GPU servers to eliminate the communication overhead?}
In the lower part of Figure~\ref{fig:feabox}, we depict an example for the proposed training framework that combines the feature extraction and the training computation within GPU servers. The intermediate I/O is eliminated by integrating the feature extraction and the training computation into a pipeline: for each batch of extracted features, we feed the batch to the model training without writing them as intermediate files into HDFS.

\vspace{0.1in}
\noindent\textbf{Challenges \& Approaches.}
However, moving the feature extraction to GPU servers is non-trivial.
Note that the number of GPU nodes is much fewer compared with the CPU-only cluster. We acknowledge two main challenges in embedding the feature extraction phase into GPU servers:

\vspace{0.05in}

\begin{enumerate}
\item {\it Network I/O bandwidth}. The network I/O bandwidth of GPU servers is by orders of magnitude smaller than the bandwidth of CPU clusters because we have fewer nodes---the total number of network adapters is lower. We materialize frequently-used features as basic features so that we can reuse them without extra I/O and computations. In addition, we use column-store that reads only the required columns in the logs to reduce I/O.
\item {\it Computing Resources}. With a smaller number of nodes, the CPU computing capability on GPU servers is also orders of magnitudes less powerful than the CPU cluster. We have to move the CPU computations to GPU operations to bridge the computing power gap.

\vspace{0.05in}

\item {\it Memory Usage}. The feature extraction process contains many memory-intensive operations, such as dictionary table lookup, sort, reduce, etc. It is desired to have an efficient memory management system to efficiently perform dynamic memory allocations on GPU servers with limited memory.
\end{enumerate}

\vspace{0.1in}

We summarize our \textbf{contributions} as follows:
\begin{itemize}
\item We propose FeatureBox, a novel end-to-end training framework that pipelines the feature extraction and the training on GPU servers.
\item We present a layer-wise operator scheduling algorithm that arranges the operators to CPUs and GPU.
\item We introduce a light-weight GPU memory management algorithm that supports dynamic GPU memory allocation with minimal overhead.
\item We experimentally evaluate FeatureBox and compare it with the previous in-production feature extraction framework on two real-world ads applications. The results confirm the effectiveness of our proposed methods.
\end{itemize}

\vspace{0.1in}
\section{Preliminary}
\vspace{0.1in}

In this section, we present a brief introduction of CTR prediction models and the hierarchical GPU parameter server. Both concepts are the foundations of FeatureBox.

\subsection{CTR Prediction Models}
About a decade ago, CTR prediction strategies with large-scale logistic regression model on carefully engineered features are proposed in~\cite{edelman2007internet,graepel2010web}. With the rapid development of deep learning, deep neural networks (DNN) attract a lot of attention in the CTR research community: The DNN model, with wide embedding layers, obtains significant improvements over classical models. The model takes a sparse high-dimensional vector as input and converts those sparse features into dense vectors through sequential embedding layers. The output dense vector is considered a low-dimensional representation of the input and is then fed into the following layers in order to compute the CTR. Most proposed CTR models share the same embedding layer architecture and only focus on the following neural network layers, see for e.g., Deep Crossing~\cite{shan2016deep}, Product-based Neural Network (PNN)~\cite{qu2016product}, Wide\&Deep Learning~\cite{cheng2016wide}, YouTube Recommendation CTR model~\cite{covington2016deep}, DeepFM~\cite{guo2017deepfm}, xDeepFM~\cite{lian2018xdeepfm} and Deep Interest Network (DIN)~\cite{zhou2018deep}. They introduce special neural layers for specific applications that capture latent feature interactions.

\begin{figure}[htbp]
\centering
\includegraphics[width=3in]{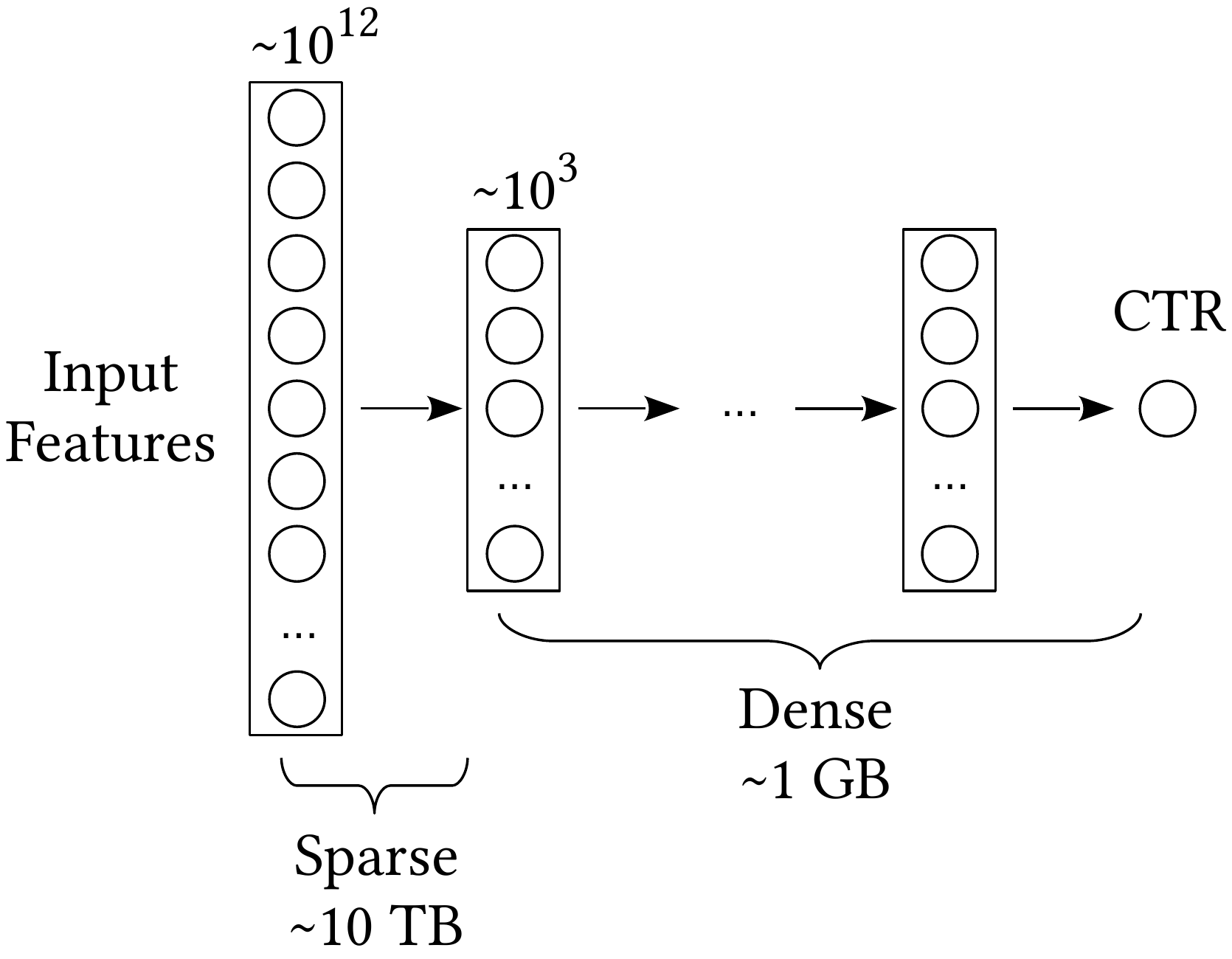}

\caption{An example for the CTR prediction network architecture.}
\label{fig:ctr}
\end{figure}

We summarize those architectures in Figure~\ref{fig:ctr}. The input features are fed to the neural network as a sparse high-dimensional vector. The dimension of the vector can be $\sim$$10^{12}$ or more. The input features for CTR models are usually from various resources with categorical values, e.g., query words, ad keywords, and user portrait. The categorical values are commonly represented as a one-hot or multi-hot encoding. Therefore, with categorical values with many sources, the number of dimensions is high ($\sim$$10^{12}$) for industry CTR prediction models. Note that, as demonstrated in~\cite{zhao2020distributed}, feature compression or hashing strategies~\cite{weinberger2009feature,li2011hashing} that reduce the number of dimensions are not fully applicable to the CTR prediction model because those solutions inevitably trade off the prediction accuracy for better computational time---recall that even a small accuracy loss leads to a noticeable online advertising revenue decrease, which is unacceptable. We embed the high-dimensional features through an embedding layer to obtain a low-dimensional ($\sim$$10^3$) representation. The number of parameters in the embedding layer can be 10 TB or more due to the high input dimension. After  the low-dimensional embedding is obtained, we fed this dense vector to the  neural network components to compute the CTR.

\subsection{Hierarchical GPU Parameter Server}
Due to the extremely high dimension of the embedding layer, the model contains more than 10 TB parameters which do not fit on most computing servers. Conventionally, the huge model is trained on an MPI cluster. We partition the model parameters across multiple computing nodes (e.g., 150 nodes) in the MPI cluster. Every computing node is assigned a batch of training data streamed directly from the HDFS.
For each node, it retrieves the required parameters from other nodes and computes the gradients for its current working mini-batch. The gradients are then updated to the nodes that maintain the corresponding parameters through MPI communications. Recently, hierarchical GPU parameter servers~\cite{zhao2020distributed} are proposed to train the massive-scale model on a limited number of GPU servers. The key observation of the hierarchical GPU parameter server is that the number of referenced parameters in a mini-batch fits the GPU memory because the input vector is sparse. It maintains three levels of hierarchical parameter servers on GPU, CPU main memory, and SSD. The working parameters are stored in GPUs, the frequently used parameters are kept in CPU main memory, and other parameters are materialized as files on SSDs. The upper-level module acts as a high-speed cache of the lower-level module. With 4 GPU nodes, the hierarchical GPU parameter server is able to be 2X faster than 150 CPU-only nodes in an MPI cluster. Our proposed FeatureBox follows the design of the training framework in the hierarchical GPU parameter server and absorbs the feature engineering workload into GPUs to eliminate excessive intermediate I/O.

\section{FeatureBox Overview}

\begin{figure}[b!]
\vspace{-0.2in}

\centering
\includegraphics[width=2.2in]{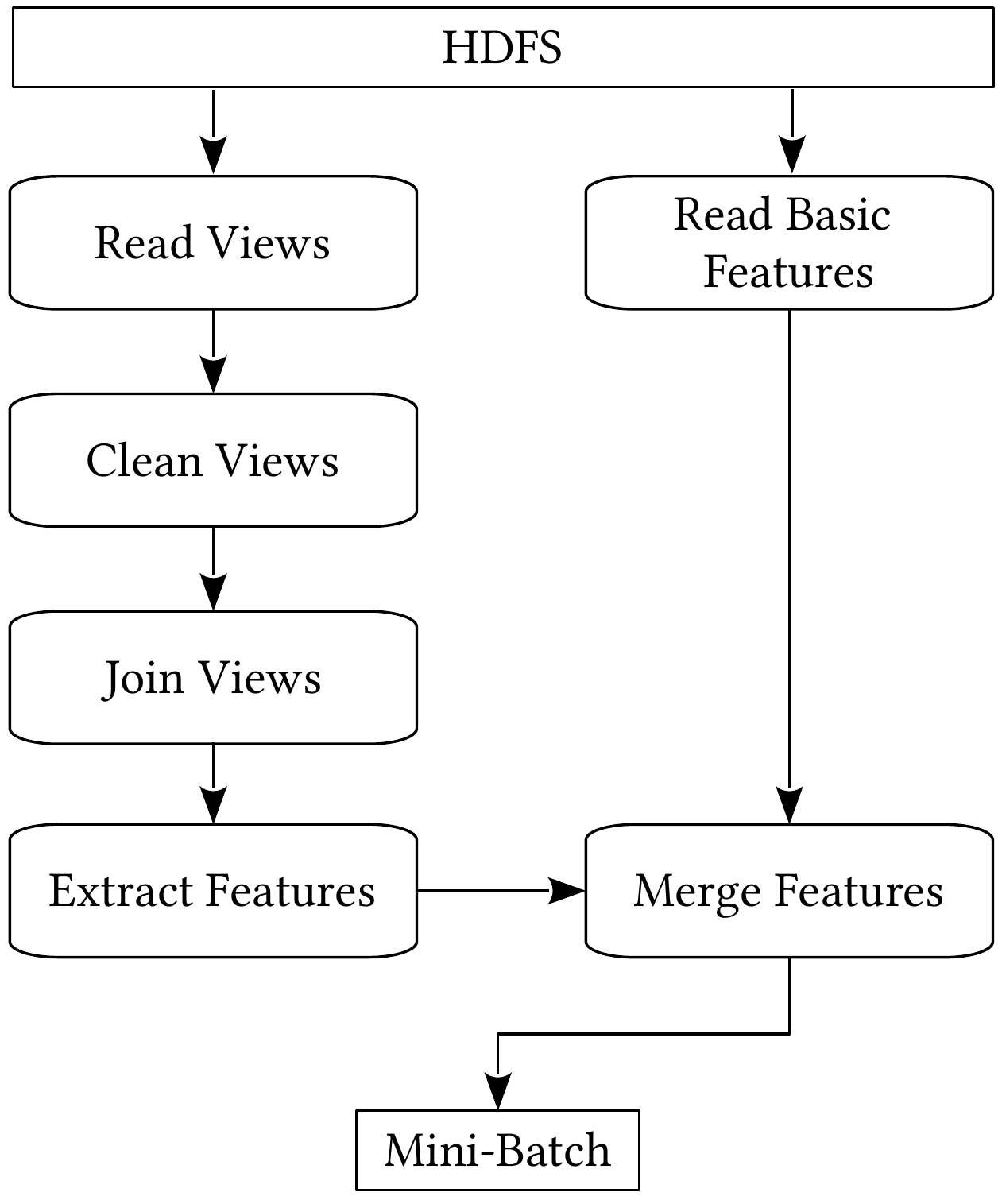}
\caption{FeatureBox pipeline.}\label{fig:pipeline}
\end{figure}

In this section, we present an overview of FeatureBox. We aim at allowing the training framework to support pipeline processing with mini-batches so that we can eliminate the excessive intermediate resulting I/O in conventional stage-after-stage methods. Figure~\ref{fig:pipeline} depicts the detailed workflow of the FeatureBox pipeline.

\begin{figure*}[t!]
\includegraphics[width=7.2in]{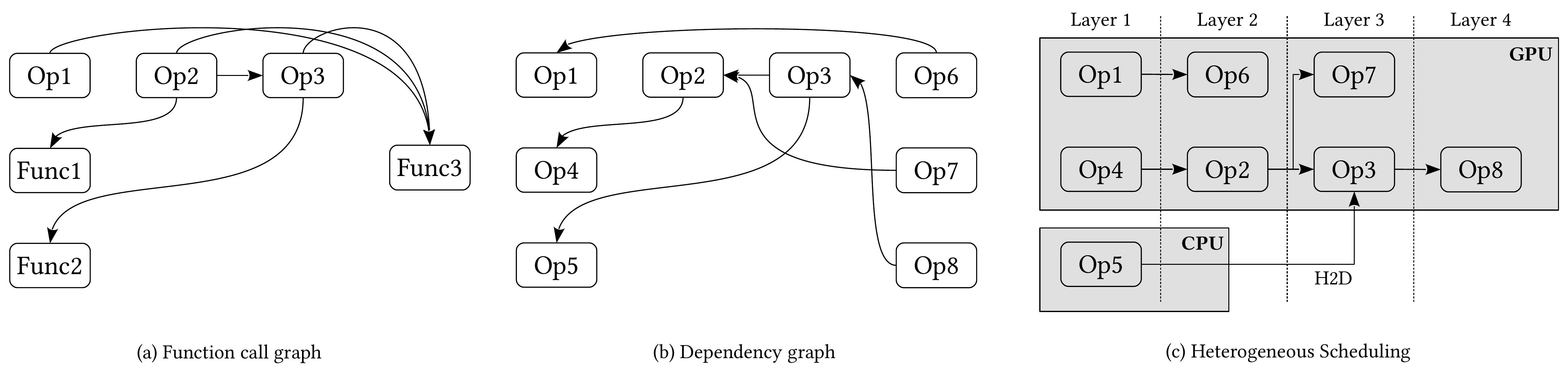}
\caption{An example for the heterogeneous operator scheduling.}
\label{fig:dag}
\end{figure*}

The workflow in Figure~\ref{fig:pipeline} has two major tracks: -- extract features from input views and -- reading basic features.
A view is a collection of raw data logs from one source, e.g., user purchase history. CTR prediction models collect features from multiple sources to obtain the best performance. The views are read from the network file system HDFS. We need to clean the views by filling null values and filtering out unrelated instances. Afterwards, the views are joined with particular keys such as user id, ads id, etc. We extract features from the joined views to obtain the desired features from the input views. Then, these features are merged with the basic features, read in a parallel path. We provide a detailed illustration for these operations as follows:

\vspace{0.1in}\noindent\textbf{Read views and basic features.}
The views and basic features are streamed from the distributed file system. The features are organized in a column-wise manner so that we only need to read the required features.

\vspace{0.1in}\noindent\textbf{Clean views.}
Views contain null values and semi-structured data, e.g., JSON format~\cite{DBLP:conf/www/PezoaRSUV16}.
At the view cleaning stage, we fill the null values and extract required fields from the semi-structured data.
Following the cleaning, all columns have non-empty and simple type (as integer, float, or string) fields. Note that the resulting views contain all the logged instances. For an application, it may not need to include all instances, e.g., an application for young people. A custom filter can be applied to filter out unrelated instances of the current application.

\vspace{0.1in}\noindent\textbf{Join views.}
We now have one structured table for each view.
Data from different views are concatenated by joining their keys, e.g., user id, ad id, etc.
We recall that the join step combines multiple views into a single structured table.

\vspace{0.1in}\noindent\textbf{Extract features}.
Every time CTR model engineers propose a new feature, an operator that computes the new feature extraction on the structured table is created.
A collection of those operators are executed in the feature extraction stage.
The FeatureBox framework figures out the dependencies of operators and schedules the execution of the operators.

\vspace{0.1in}\noindent\textbf{Merge features}.
The extracted features are further merged with the basic features read from HDFS. The merging is also realized by a join operation on the instance id, which is a unique value generated when an instance is logged.
Subsequent to the merging, a mini-batch of training data is generated and is fed to the neural network for the training.

\vspace{0.1in}
\section{Heterogeneous Operator Scheduling}
\vspace{0.1in}

The stages discussed above are represented as operators in the FeatureBox pipeline. Note that those operators are heterogeneous: Some operators are network I/O intensive, e.g., read views and read basic features; some operators are computation-intensive, e.g., clean views and extract features; and the remaining operators with joining, e.g., join views and merge features, rely on heavy memory consumption for large table joins (which corresponds to a large dictionary lookup). Therefore, we introduce a heterogeneous operator scheduler that manages the operator execution on both CPUs and GPUs.

\vspace{0.1in}\noindent\textbf{Scheduling.}
Figure~\ref{fig:dag} shows an example for the heterogeneous operator scheduling algorithm. We first present a function call graph for operators in Figure~\ref{fig:dag}(a).
Three operators and three major functions are displayed in the example.
Op1 calls Func3; Op2 calls Func1 and Func3; and Op3 calls Func2 and Func3, where Func1 and Func2 are pre-processing calls, and Func3 is a post-processing call. We make a fine granularity pipeline so that the initialing overhead of the pipeline is minimized. The fine-granularity is obtained by viewing each function call as a separate operator. Then, we obtain 5 more operators: Op4 is a call for Func1; Op5 is a call for Func2; Op6, Op7, and Op8 are the Func3 calls from Op1, Op2, and Op3, respectively. Their dependency graph is illustrated in Figure~\ref{fig:dag}(b). Now we have a directed acyclic graph (DAG) for the operators. As shown in Figure~\ref{fig:dag}(c), we perform a topological sort on the dependency graph, assign the operators with no dependencies (root operators) to the first layer, and put the remaining operators to the corresponding layer according to their depth from the root operators. With this layer-wise partition, we observe that the operators in the same layer do not have any execution dependency. We issue the operators in the same layer together and perform a synchronization at the end of each layer to ensure the execution dependency. We prefer to execute operators on GPUs unless an operator requires a significant memory footprint that does not fit in the GPU memory.
For instance, Op5 (Func2) in Figure~\ref{fig:dag} is a word embedding table look up operation that requires a considerable amount of memory. We assign this operation to CPU workers and move its results from the CPU main memory to GPUs as a host-to-device (H2D) CUDA call.

\vspace{0.1in}\noindent\textbf{Inner-GPU operator launching.}
After the layer-wise DAG operator scheduling, we have determined the execution device for each operator and the synchronization barriers. However, CUDA kernel launching is has a noticeable overhead.
We report the CUDA kernel launch overhead in Table~\ref{tbl:launch}.

\begin{table}[htbp]
\caption{The kernel launching overhead with an empty kernel on Nvidia Tesla V100-SXM2-32GB.}
\label{tbl:launch}
\centering
\normalsize
\begin{tabular}{c|rrrrr}
\toprule
\hline
\#Launches & 1 & 10 & 100 & 1,000 & 10,000 \\
\hline
Time (us) & 4 & 35 & 360 & 3,619 & 34,515\\
\hline
\bottomrule
\end{tabular}\vspace{0.15in}
\end{table}

The test is performed on an Nvidia Tesla V100-SXM2-32GB GPU for an empty kernel with 5 pointer-type arguments. The CUDA driver version is 10.2. The average launching time for a kernel is around 3.5 us. Since we have fine-granularity operators, we have to rapidly launch CUDA kernels to execute the large number of operators. In order to eliminate the launching overhead, we rewrite the operator kernel as a CUDA device function for each operator in the same layer and create a meta-kernel that sequentially executes the operator device functions in a runtime-compilation manner. The overhead of the meta-kernel generation is disregarded---we only need to create this meta-kernel for each layer once as a pre-processing of the training since we determine the operator execution order before the actual training phase and keep the scheduling fixed.
With the generated meta-kernels, we only need to launch one kernel for each layer.

\vspace{0.15in}
\section{GPU Memory Management}
\vspace{0.15in}

Feature extraction operators usually need to cope with strings of varying length, e.g., query keywords and ads titles. The execution of the operator commonly dynamically allocates memory to process the strings. For example, splitting a string with a delimiter needs to allocate an array to store the result of the splitting operation. We propose a light-weight block-level GPU memory pool to accelerate this dynamic allocation.

\begin{figure}[htbp]
\centering
\includegraphics[width=.5\textwidth]{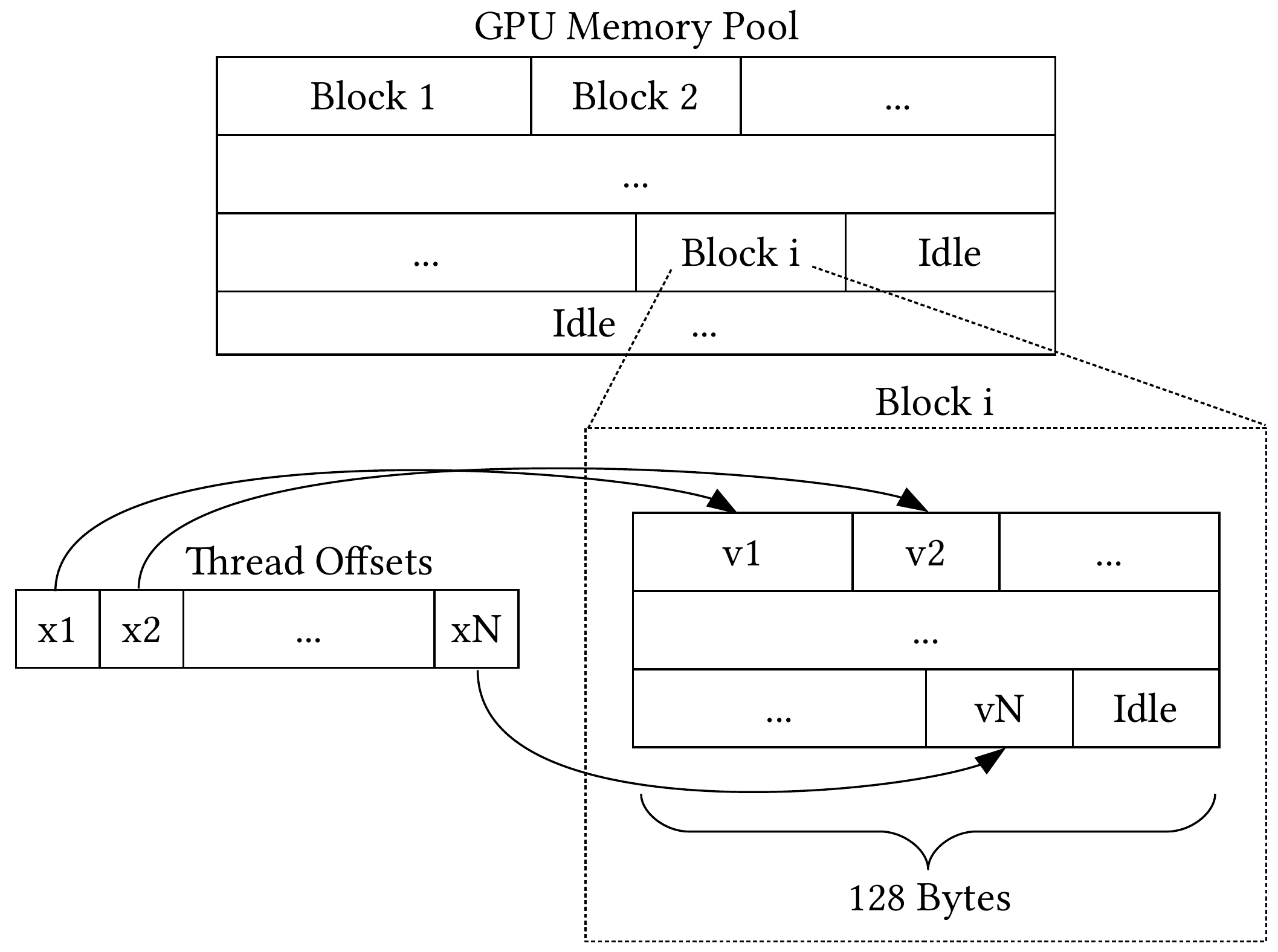}
\caption{A visual illustration for the GPU memory pool architecture.}
\label{fig:memory}
\end{figure}

Figure~\ref{fig:memory} presents a visual illustration for our proposed block-level GPU memory pool.
The \texttt{Thread Offsets} denotes an array that stores the pointers to the dynamically allocated memory in the GPU memory pool. The memory in the GPU memory pool is pre-allocated in the GPU global memory. For each block, the allocated memory is aligned in 128 bytes for a cache-friendly execution.

\vspace{0.1in}\noindent\textbf{Dynamic GPU memory allocation.} Algorithm~\ref{alg:allocate} describes the workflow of the in-kernel dynamic memory allocation. We maintain a global variable \textit{idle\_memory\_head} that stores the pointer of the head address of our pre-allocated GPU memory pool. We assume each GPU thread in a block has computed their required allocation size $\textit{size}_i$. We first compute an in-block parallel prefix sum on $\textit{size}_{1..N}$ to obtain the prefix sum $\textit{prefix}_{1..N}$, where $N$ is the number of threads in a block.
The prefix sum is used to compute the total size of the requested memory. In addition, we can easily compute the thread offsets by adding the prefix sum to the head of the allocated memory address.
After that, we let one thread in the block, e.g., thread~1, to apply the memory for the entire block---the total size is $\textit{prefix}_N$.
The memory allocation is implemented by an \textit{atomic\_add} operation. Line~\ref{line:atomic} calls the CUDA atomic add that adds $\textit{prefix}_N$ to \textit{idle\_memory\_head} and returns the old value of \textit{idle\_memory\_head} to \textit{address} in an atomic fashion---no data race within this operation.
Once the requested memory is allocated for the block, we increment the \textit{idle\_memory\_head} pointer in the memory pool.
We finalize the allocation by letting all threads in the block compute their corresponding offsets by adding the prefix sum to the allocated address.
The memory allocation is called inside the meta-kernel that we generated in the operator scheduling.
The entire allocation process has very little overhead costs---it does not require any inter-block synchronization or any kernel launches.

\begin{algorithm}[b!]
\caption{In-Kernel Dynamic Memory Allocation}\label{alg:allocate}
\textbf{Input:} allocation memory size for the $i^\textit{th}$ thread, $\textit{size}_i$; global memory pool head pointer, $\textit{idle\_memory\_head}$;\\
\textbf{Output:} thread offsets, $\textit{offsets}_i$;
\begin{algorithmic}[1]
\STATE $\textit{prefix}_{1..N} \leftarrow \textit{parallel\_prefix\_sum}(\textit{size}_{1..N})$
\STATE $\textit{address} \leftarrow \textit{atomic\_add}(\textit{idle\_memory\_head},\textit{prefix}_N)$\label{line:atomic}
\FOR{\textbf{each} thread $i$ in the current block \textbf{concurrently}}
    \STATE $\textit{offsets}_i \leftarrow \textit{address} + \textit{prefix}_i- \textit{prefix}_1$
\ENDFOR
\end{algorithmic}
\end{algorithm}

\begin{table*}[t]
\caption{End-to-end training of MapReduce feature extraction with hierarchical GPU parameter server and FeatureBox.}\label{tbl:comp}
\normalsize
\centering
\begin{tabular}{l|c|c|c|c}
\toprule
\hline
 & \multicolumn{2}{c|}{Application A} & \multicolumn{2}{c}{Application B} \\
 \hline
\#Instances & \multicolumn{2}{c|}{$\sim 1\times10^9$} & \multicolumn{2}{c}{$\sim 2\times 10^9$} \\
Log Size & \multicolumn{2}{c|}{$\sim$15 TB} & \multicolumn{2}{c}{$\sim$25 TB} \\
\hline
Framework & MapReduce + GPU & \multicolumn{1}{l|}{FeatureBox} & MapReduce + GPU & \multicolumn{1}{l}{FeatureBox} \\
\hline
\#Machines & 20 CPU + 1 GPU & 1 GPU & 30 CPU + 2 GPU & 2 GPU \\
Execution Time & \multicolumn{1}{c|}{18h} & 3.5h & \multicolumn{1}{c|}{27h} & 2.65h \\
Speedup & \multicolumn{1}{c|}{-} & 5.14X & \multicolumn{1}{c|}{-} & 10.19X\\
Intermediate I/O Saving & - & $\sim$50 TB & - & $\sim$100 TB\\
\hline
\bottomrule
\end{tabular}
\end{table*}

\newpage
\vspace{0.1in}\noindent\textbf{Reset GPU memory pool.}
Our light-weight memory allocation strategy only maintains a pointer on a pre-allocated continuous global memory. However, the single-pointer design does not support memory freeing.
We have to maintain an additional collection of freed memory and allocate the requested memory chunks from this collection---the maintenance of this additional data structure leads to significant memory allocation overhead.
We observe that our operators are in fine-granularity and are scheduled layer by layer. Therefore, we can assume that the total required memory for dynamic allocations fits the GPU memory. We perform the memory release in a batch fashion: the memory pool is reset after each meta-kernel. The reset can be done in a constant time---we only need to set \textit{idle\_memory\_head} to the original allocated memory address for the memory pool so that the allocation request in the meta-kernel for the following layer gets the allocation from the beginning of the memory pool.

\vspace{0.1in}
\section{Experimental Evaluation}
\vspace{0.1in}

In this section, we investigate the effectiveness of our proposed framework FeatureBox through a set of numerical experiments.
Specifically, the experiments are targeted to address the following questions:
\begin{itemize}
\item How is the end-to-end training time of FeatureBox compared with the previous MapReduce solution?
\item How much intermediate I/O is saved by the pipelining architecture?
\item What is the performance of FeatureBox in the feature extraction task?
\end{itemize}

\vspace{0.1in}\noindent\textbf{Systems.}
The MapReduce feature extraction baseline is our previous in-production solution to extract features for the training tasks. It runs in an MPI cluster with CPU-only nodes in a data center. Commonly, a feature extraction job requires 20 to 30 nodes. Each node is equipped with server-grade CPUs ($\sim$100 threads). The training part is executed on GPU nodes. Each GPU node has 8 cutting-edge 32 GB HBM GPUs, $\sim$1 TB main memory, $\sim$20 TB RAID-0 NVMe SSDs, and a 100 Gb RDMA network adaptor. The training framework is the hierarchical GPU parameter server.
All nodes are inter-connected through a high-speed Ethernet switch.

\vspace{0.1in}\noindent\textbf{Models.}
We use CTR prediction models on two real-world online advertising applications. The neural network backbones of both models follow the design in Figure~\ref{fig:ctr}. The major difference between the two models is the number of input features. Both models have more than $\sim$10 TB parameters. We collect real user click history logs as the training dataset.

\subsection{End-to-End Training }
We report Table~\ref{tbl:comp} specifications about the training data and the end-to-end training comparison between our proposed FeatureBox and the MapReduce feature extraction with hierarchical GPU parameter server training as a baseline. Both training datasets contain billions of instances. The size of the logs is $\sim$15 TB for application A, and $\sim$25 TB for application B. The end-to-end training time includes the features extraction from the log time and the model training time. FeatureBox uses 1 GPU server for application A and 2 GPU servers for application B. In addition to the GPU servers, the baseline solution also employs 20/30 CPU-only servers to perform feature extraction. The baseline solution first extracts features using MapReduce, saves the features as training data in HDFS, and streams the generated training data to the GPU servers to train the model. On the other hand, FeatureBox processes the data in a pipeline fashion: features are extracted on GPU servers and then are immediately fed to the training framework on the same GPU server. For application A, FeatureBox only takes 3.5 hours to finish the feature extraction and the training while the baseline solution requires 18 hours---with fewer number of machines, FeatureBox has a \emph{5.14X} speedup compared to the baseline.
Meanwhile, Application B presents a bigger volume of log instances. Hence, we use two GPU servers to perform the training. We can observe a larger gap between FeatureBox and the baseline when the data size scales up: FeatureBox outperforms the baseline with a \emph{10.19X} speedup. One of the main reasons of the speedup is that FeatureBox eliminates the huge intermediate I/O from the MapReduce framework.
We save $\sim$50-100 TB intermediate I/O while using FeatureBox.

\subsection{Feature Extraction}
Although the improvement of FeatureBox in the end-to-end training time mainly benefits from the pipeline architecture, we also investigate the feature extraction performance to confirm that our proposed GPU feature extraction framework is a better alternative to the baseline MapReduce solution.

\begin{figure}[htbp]
\centering
\includegraphics[width=3in]{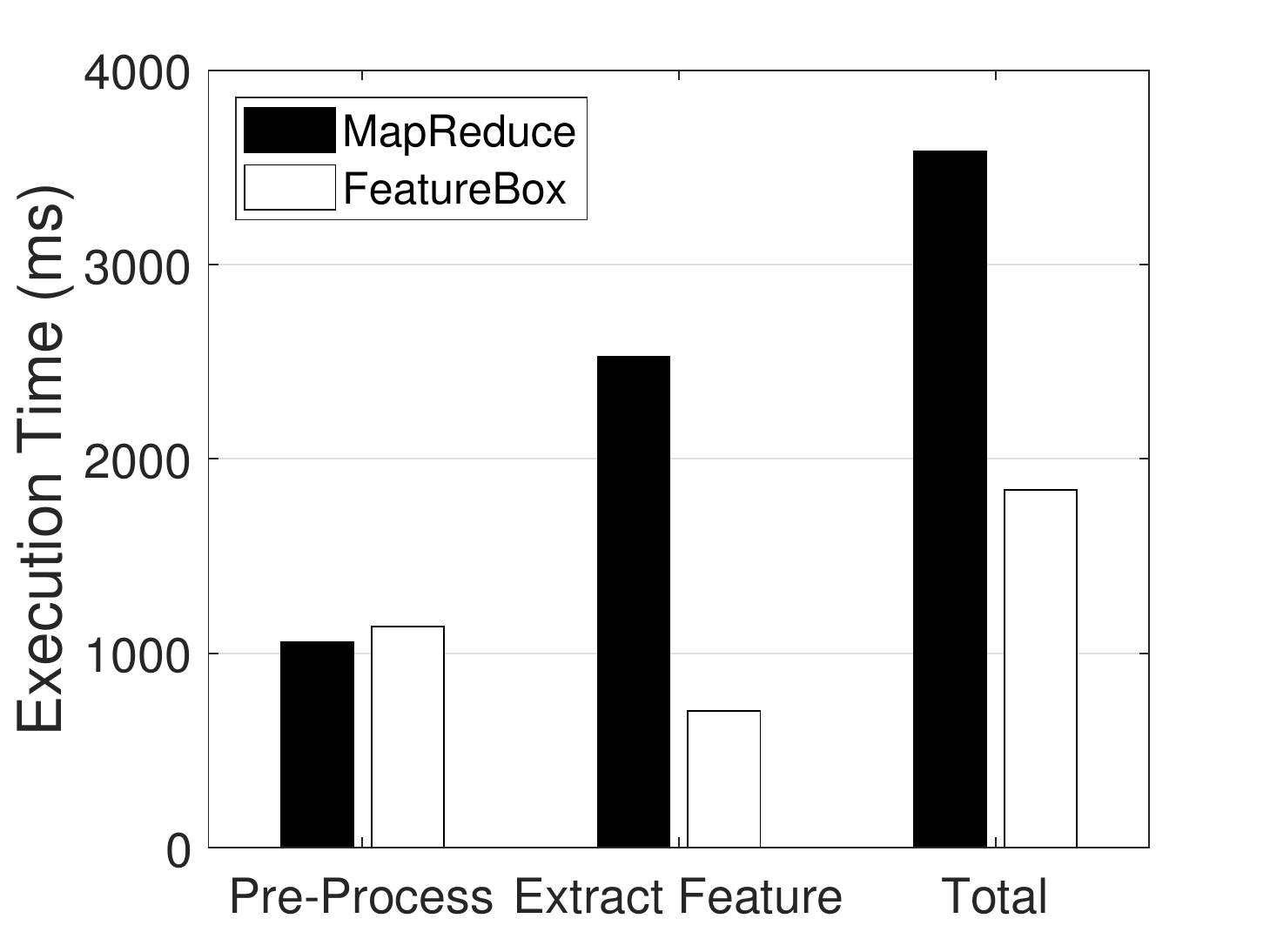}
\caption{Feature extraction time of MapReduce and FeatureBox.}
\label{fig:extract}
\end{figure}

We report, in Figure~\ref{fig:extract}, the time to extract features from $10,000$ log instances of Application B. MapReduce runs on 30 CPU-only servers and FeatureBox runs on 2 GPU servers. The pre-processing time includes the stages to prepare the data for the feature extraction, such as read, clean, and join views. The pre-processing time of both methods are comparable because the executed operations are mostly memory and network I/O. Regarding the time to extract features, FeatureBox is more than 3 times faster than MapReduce. FeatureBox only takes around half of the time to extract the features than the baseline.

\subsection{Discussion}
Based on these results, we can answer the questions that drive the experiments: The end-to-end training time of FeatureBox is 5-10 times faster than the baseline. Due to the pipeline design, FeatureBox saves us 50-100 TB intermediate I/O. For feature extraction only tasks, FeatureBox on 2 GPU servers is 2X faster than MapReduce on 30 CPU-only servers.

\vspace{0.1in}
\section{Conclusions}
\vspace{0.1in}

In this paper, we introduce FeatureBox, a novel end-to-end training framework that pipelines the feature extraction and the training on GPU servers to save the intermediate I/O of the feature extraction. We rewrite computation-intensive feature extraction operators as GPU operators and leave the memory-intensive operator on CPUs. We introduce a layer-wise operator scheduling algorithm to schedule these heterogeneous operators. We present a light-weight GPU memory management algorithm that supports dynamic GPU memory allocation with minimal overhead. We experimentally evaluate FeatureBox and compare it with the previous in-production MapReduce feature extraction framework on two real-world ads applications. The results show that FeatureBox is 5-10X faster than the baseline.

\bibliographystyle{plain}
\bibliography{refs_scholar}

\begin{thebibliography}{10}

\bibitem{borthakur2008hdfs}
Dhruba Borthakur.
\newblock {HDFS} architecture guide.
\newblock {\em Hadoop apache project}, 53(1-13):2, 2008.

\bibitem{cheng2016wide}
Heng{-}Tze Cheng, Levent Koc, Jeremiah Harmsen, Tal Shaked, Tushar Chandra,
  Hrishi Aradhye, Glen Anderson, Greg Corrado, Wei Chai, Mustafa Ispir, Rohan
  Anil, Zakaria Haque, Lichan Hong, Vihan Jain, Xiaobing Liu, and Hemal Shah.
\newblock Wide \& deep learning for recommender systems.
\newblock In {\em Proceedings of the 1st Workshop on Deep Learning for
  Recommender Systems, (DLRS@RecSys)}, pages 7--10, Boston, MA, 2016.

\bibitem{covington2016deep}
Paul Covington, Jay Adams, and Emre Sargin.
\newblock Deep neural networks for youtube recommendations.
\newblock In {\em Proceedings of the 10th {ACM} Conference on Recommender
  Systems (RecSys)}, pages 191--198, Boston, MA, 2016.

\bibitem{dean2008mapreduce}
Jeffrey Dean and Sanjay Ghemawat.
\newblock Mapreduce: simplified data processing on large clusters.
\newblock {\em Communications of the ACM}, 51(1):107--113, 2008.

\bibitem{dosovitskiy2021image}
Alexey Dosovitskiy, Lucas Beyer, Alexander Kolesnikov, Dirk Weissenborn,
  Xiaohua Zhai, Thomas Unterthiner, Mostafa Dehghani, Matthias Minderer, Georg
  Heigold, Sylvain Gelly, Jakob Uszkoreit, and Neil Houlsby.
\newblock An image is worth 16x16 words: Transformers for image recognition at
  scale.
\newblock In {\em 9th International Conference on Learning Representations
  (ICLR)}, Virtual Event, Austria, 2021.

\bibitem{edelman2007internet}
Benjamin Edelman, Michael Ostrovsky, and Michael Schwarz.
\newblock Internet advertising and the generalized second-price auction:
  Selling billions of dollars worth of keywords.
\newblock {\em American economic review}, 97(1):242--259, 2007.

\bibitem{fan2019mobius}
Miao Fan, Jiacheng Guo, Shuai Zhu, Shuo Miao, Mingming Sun, and Ping Li.
\newblock {MOBIUS:} towards the next generation of query-ad matching in baidu's
  sponsored search.
\newblock In {\em Proceedings of the 25th {ACM} {SIGKDD} International
  Conference on Knowledge Discovery {\&} Data Mining (KDD)}, pages 2509--2517,
  Anchorage, AK, 2019.

\bibitem{fei2019hierarchical}
Hongliang Fei, Shulong Tan, and Ping Li.
\newblock Hierarchical multi-task word embedding learning for synonym
  prediction.
\newblock In {\em Proceedings of the 25th {ACM} {SIGKDD} International
  Conference on Knowledge Discovery {\&} Data Mining (KDD)}, pages 834--842,
  Anchorage, AK, 2019.

\bibitem{fei2021gemnn}
Hongliang Fei, Jingyuan Zhang, Xingxuan Zhou, Junhao Zhao, Xinyang Qi, and Ping
  Li.
\newblock {GemNN}: Gating-enhanced multi-task neural networks with feature
  interaction learning for {CTR} prediction.
\newblock In {\em Proceedings of the 44th International {ACM} {SIGIR}
  Conference on Research and Development in Information Retrieval (SIGIR)},
  pages 2166--2171, Virtual Event, Canada, 2021.

\bibitem{goodfellow2014generative}
Ian~J. Goodfellow, Jean Pouget{-}Abadie, Mehdi Mirza, Bing Xu, David
  Warde{-}Farley, Sherjil Ozair, Aaron~C. Courville, and Yoshua Bengio.
\newblock Generative adversarial nets.
\newblock In {\em Advances in Neural Information Processing Systems (NIPS)},
  pages 2672--2680, Montreal, Canada, 2014.

\bibitem{graepel2010web}
Thore Graepel, Joaquin~Qui{\~{n}}onero Candela, Thomas Borchert, and Ralf
  Herbrich.
\newblock Web-scale bayesian click-through rate prediction for sponsored search
  advertising in microsoft's bing search engine.
\newblock In {\em Proceedings of the 27th International Conference on Machine
  Learning (ICML)}, pages 13--20, Haifa, Israel, 2010.

\bibitem{guo2017deepfm}
Huifeng Guo, Ruiming Tang, Yunming Ye, Zhenguo Li, and Xiuqiang He.
\newblock {DeepFM}: {A} factorization-machine based neural network for {CTR}
  prediction.
\newblock In {\em Proceedings of the Twenty-Sixth International Joint
  Conference on Artificial Intelligence (IJCAI)}, pages 1725--1731, Melbourne,
  Australia, 2017.

\bibitem{he2016deep}
Kaiming He, Xiangyu Zhang, Shaoqing Ren, and Jian Sun.
\newblock Deep residual learning for image recognition.
\newblock In {\em Proceedings of the 2016 {IEEE} Conference on Computer Vision
  and Pattern Recognition (CVPR)}, pages 770--778, Las Vegas, NV, 2016.

\bibitem{kang2020novel}
Tianyu Kang, Ping Chen, John Quackenbush, and Wei Ding.
\newblock A novel deep learning model by stacking conditional restricted
  boltzmann machine and deep neural network.
\newblock In {\em Proceedings of the 26th {ACM} {SIGKDD} Conference on
  Knowledge Discovery and Data Mining (KDD)}, pages 1316--1324, Virtual Event,
  CA, 2020.

\bibitem{li2020video}
Dingcheng Li, Xu~Li, Jun Wang, and Ping Li.
\newblock Video recommendation with multi-gate mixture of experts soft actor
  critic.
\newblock In {\em Proceedings of the 43rd International {ACM} {SIGIR}
  conference on research and development in Information Retrieval (SIGIR)},
  pages 1553--1556, Virtual Event, China, 2020.

\bibitem{li2011hashing}
Ping Li, Anshumali Shrivastava, Joshua~L. Moore, and Arnd~Christian
  K{\"{o}}nig.
\newblock Hashing algorithms for large-scale learning.
\newblock In {\em Advances in Neural Information Processing Systems (NIPS)},
  pages 2672--2680, Granada, Spain, 2011.

\bibitem{li2021learning}
Xijun Li, Weilin Luo, Mingxuan Yuan, Jun Wang, Jiawen Lu, Jie Wang, Jinhu
  L{\"{u}}, and Jia Zeng.
\newblock Learning to optimize industry-scale dynamic pickup and delivery
  problems.
\newblock In {\em Proceedings of the 37th {IEEE} International Conference on
  Data Engineering (ICDE)}, pages 2511--2522, Chania, Greece, 2021.

\bibitem{lian2018xdeepfm}
Jianxun Lian, Xiaohuan Zhou, Fuzheng Zhang, Zhongxia Chen, Xing Xie, and
  Guangzhong Sun.
\newblock xdeepfm: Combining explicit and implicit feature interactions for
  recommender systems.
\newblock In {\em Proceedings of the 24th ACM SIGKDD International Conference
  on Knowledge Discovery {\&} Data Mining (KDD)}, pages 1754--1763, London, UK,
  2018.

\bibitem{liu2020autogroup}
Bin Liu, Niannan Xue, Huifeng Guo, Ruiming Tang, Stefanos Zafeiriou, Xiuqiang
  He, and Zhenguo Li.
\newblock Autogroup: Automatic feature grouping for modelling explicit
  high-order feature interactions in {CTR} prediction.
\newblock In {\em Proceedings of the 43rd International {ACM} {SIGIR}
  conference on research and development in Information Retrieval (SIGIR)},
  pages 199--208, Virtual Event, China, 2020.

\bibitem{ma2020temporal}
Yifei Ma, Balakrishnan~(Murali) Narayanaswamy, Haibin Lin, and Hao Ding.
\newblock Temporal-contextual recommendation in real-time.
\newblock In {\em Proceedings of the 26th {ACM} {SIGKDD} Conference on
  Knowledge Discovery and Data Mining (KDD)}, pages 2291--2299, Virtual Event,
  CA, 2020.

\bibitem{merhej2021what}
Charbel Merhej, Ryan~J. Beal, Tim Matthews, and Sarvapali~D. Ramchurn.
\newblock What happened next? using deep learning to value defensive actions in
  football event-data.
\newblock In {\em Proceedings of the 27th {ACM} {SIGKDD} Conference on
  Knowledge Discovery and Data Mining (KDD)}, pages 3394--3403, Virtual Event,
  Singapore, 2021.

\bibitem{nguyen2019machine}
Giang Nguyen, Stefan Dlugolinsky, Martin Bob{\'a}k, Viet Tran,
  {\'A}lvaro~L{\'o}pez Garc{\'\i}a, Ignacio Heredia, Peter Mal{\'\i}k, and
  Ladislav Hluch{\`y}.
\newblock Machine learning and deep learning frameworks and libraries for
  large-scale data mining: a survey.
\newblock {\em Artificial Intelligence Review}, 52(1):77--124, 2019.

\bibitem{DBLP:conf/www/PezoaRSUV16}
Felipe Pezoa, Juan~L. Reutter, Fernando Su{\'{a}}rez, Mart{\'{\i}}n Ugarte, and
  Domagoj Vrgoc.
\newblock Foundations of {JSON} schema.
\newblock In {\em Proceedings of the 25th International Conference on World
  Wide Web (WWW)}, pages 263--273, Montreal, Canada, 2016.

\bibitem{qi2021pp}
Tao Qi, Fangzhao Wu, Chuhan Wu, and Yongfeng Huang.
\newblock {PP-Rec}: News recommendation with personalized user interest and
  time-aware news popularity.
\newblock In {\em Proceedings of the 59th Annual Meeting of the Association for
  Computational Linguistics and the 11th International Joint Conference on
  Natural Language Processing (ACL)}, pages 5457--5467, Virtual Event, 2021.

\bibitem{qu2016product}
Yanru Qu, Han Cai, Kan Ren, Weinan Zhang, Yong Yu, Ying Wen, and Jun Wang.
\newblock Product-based neural networks for user response prediction.
\newblock In {\em Proceedings of the 2016 IEEE 16th International Conference on
  Data Mining (ICDM)}, pages 1149--1154, Barcelona, Spain, 2016.

\bibitem{shan2016deep}
Ying Shan, T.~Ryan Hoens, Jian Jiao, Haijing Wang, Dong Yu, and J.~C. Mao.
\newblock Deep crossing: Web-scale modeling without manually crafted
  combinatorial features.
\newblock In {\em Proceedings of the 22nd ACM SIGKDD International Conference
  on Knowledge Discovery and Data Mining (KDD)}, pages 255--262, San Francisco,
  CA, 2016.

\bibitem{sheng2021one}
Xiang{-}Rong Sheng, Liqin Zhao, Guorui Zhou, Xinyao Ding, Binding Dai, Qiang
  Luo, Siran Yang, Jingshan Lv, Chi Zhang, Hongbo Deng, and Xiaoqiang Zhu.
\newblock One model to serve all: Star topology adaptive recommender for
  multi-domain {CTR} prediction.
\newblock In {\em Proceedings of the 30th {ACM} International Conference on
  Information and Knowledge Management (CIKM)}, pages 4104--4113, Virtual
  Event, Queensland, Australia, 2021.

\bibitem{voulodimos2018deep}
Athanasios Voulodimos, Nikolaos Doulamis, Anastasios~D. Doulamis, and Eftychios
  Protopapadakis.
\newblock Deep learning for computer vision: {A} brief review.
\newblock {\em Comput. Intell. Neurosci.}, 2018:7068349:1--7068349:13, 2018.

\bibitem{wei2017collaborative}
Jian Wei, Jianhua He, Kai Chen, Yi~Zhou, and Zuoyin Tang.
\newblock Collaborative filtering and deep learning based recommendation system
  for cold start items.
\newblock {\em Expert Systems with Applications}, 69:29--39, 2017.

\bibitem{weinberger2009feature}
Kilian~Q. Weinberger, Anirban Dasgupta, John Langford, Alexander~J. Smola, and
  Josh Attenberg.
\newblock Feature hashing for large scale multitask learning.
\newblock In {\em Proceedings of the 26th Annual International Conference on
  Machine Learning (ICML)}, pages 1113--1120, Montreal, Canada, 2009.

\bibitem{yu2022boost}
Tan Yu, Zhipeng Jin, Jie Liu, Yi~Yang, Hongliang Fei, and Ping Li.
\newblock Boost ctr prediction for new advertisements via modeling visual
  content.
\newblock {\em arXiv preprint arXiv:2209.11727}, 2022.

\bibitem{zamany2022towards}
Siamak Zamany, Dingcheng Li, Hongliang Fei, and Ping Li.
\newblock Towards deeper understanding of variational auto-encoders for binary
  collaborative filtering.
\newblock In {\em Proceedings of the 2022 {ACM} {SIGIR} International
  Conference on the Theory of Information Retrieval (ICTIR)}, pages 254--263,
  Madrid, Spain, 2022.

\bibitem{zerveas2021transformer}
George Zerveas, Srideepika Jayaraman, Dhaval Patel, Anuradha Bhamidipaty, and
  Carsten Eickhoff.
\newblock A transformer-based framework for multivariate time series
  representation learning.
\newblock In {\em Proceedings of the 27th {ACM} {SIGKDD} Conference on
  Knowledge Discovery and Data Mining (kDD)}, pages 2114--2124, Virtual Event,
  Singapore, 2021.

\bibitem{zhang2021multi}
Kai Zhang, Hao Qian, Qing Cui, Qi~Liu, Longfei Li, Jun Zhou, Jianhui Ma, and
  Enhong Chen.
\newblock Multi-interactive attention network for fine-grained feature learning
  in {CTR} prediction.
\newblock In {\em Proceedings of the Fourteenth {ACM} International Conference
  on Web Search and Data Mining (WSDM)}, pages 984--992, Virtual Event, Israel,
  2021.

\bibitem{zhang2019deep}
Shuai Zhang, Lina Yao, Aixin Sun, and Yi~Tay.
\newblock Deep learning based recommender system: A survey and new
  perspectives.
\newblock {\em ACM Computing Surveys (CSUR)}, 52(1):1--38, 2019.

\bibitem{zhao2021rlnf}
Pu~Zhao, Chuan Luo, Cheng Zhou, Bo~Qiao, Jiale He, Liangjie Zhang, and Qingwei
  Lin.
\newblock {RLNF:} reinforcement learning based noise filtering for
  click-through rate prediction.
\newblock In {\em Proceedings of the 44th International {ACM} {SIGIR}
  Conference on Research and Development in Information Retrieval (SIGIR)},
  pages 2268--2272, Virtual Event, Canada, 2021.

\bibitem{zhao2020distributed}
Weijie Zhao, Deping Xie, Ronglai Jia, Yulei Qian, Ruiquan Ding, Mingming Sun,
  and Ping Li.
\newblock Distributed hierarchical {GPU} parameter server for massive scale
  deep learning ads systems.
\newblock In {\em Proceedings of Machine Learning and Systems 2020 (MLSys)},
  Austin, TX, 2020.

\bibitem{zhao2019aibox}
Weijie Zhao, Jingyuan Zhang, Deping Xie, Yulei Qian, Ronglai Jia, and Ping Li.
\newblock {AIBox}: {CTR} prediction model training on a single node.
\newblock In {\em Proceedings of the 28th {ACM} International Conference on
  Information and Knowledge Management (CIKM)}, pages 319--328, Beijing, China,
  2019.

\bibitem{zhou2018deep}
Guorui Zhou, Xiaoqiang Zhu, Chengru Song, Ying Fan, Han Zhu, Xiao Ma, Yanghui
  Yan, Junqi Jin, Han Li, and Kun Gai.
\newblock Deep interest network for click-through rate prediction.
\newblock In {\em Proceedings of the 24th ACM SIGKDD International Conference
  on Knowledge Discovery \& Data Mining (KDD)}, pages 1059--1068, London, UK,
  2018.

\bibitem{zhu2020emsembled}
Jieming Zhu, Jinyang Liu, Weiqi Li, Jincai Lai, Xiuqiang He, Liang Chen, and
  Zibin Zheng.
\newblock Ensembled {CTR} prediction via knowledge distillation.
\newblock In {\em Proceedings of the 29th {ACM} International Conference on
  Information and Knowledge Management (CIKM)}, pages 2941--2958, Virtual
  Event, Ireland, 2020.

\end{thebibliography}

\end{document}